      [1999/12/01 v1.4c Il Nuovo Cimento]
\documentclass[varenna]{cimento}
\usepackage{graphicx}  
\title{Gauge and Lorentz invariant pionic correlations in quasi-elastic 
electron scattering}
\author{M.B.~Barbaro}
\institute{Dipartimento di Fisica Teorica and INFN - 
Universit\'a di Torino, Via P.~Giuria1, I- 10125 Turin, Italy 
}

\newcommand{\Kbar}{\not{\!K}}

\newcommand{\be}{\begin{equation}}
\newcommand{\ee}{\end{equation}}
\newcommand{\ba}{\begin{eqnarray}}
\newcommand{\ea}{\end{eqnarray}}

\newcommand{\nh}{{\bf      h}}

\newcommand{\nk}{{\bf      k}}
\newcommand{\np}{{\bf      p}}       
\newcommand{\nq}{{\bf      q}}

\begin{document}

\maketitle

\newpage

\section{Introduction}

The problem of accounting for relativistic dynamics in nuclear physics 
is a daunting one and far from being solved (see, for example, 
Refs.~\cite{Ser86,Cel86}).  
However in modern experimental studies of electron scattering from 
nuclei the typical values of energy and momentum
transfer are comparable to, or even larger than, the scale set by
the nucleon mass: accordingly one must expect relativistic effects
to be important. 

One reason for going to high energy and momentum transfers is the 
possibility
of extracting information on the nucleon's form factors, in particular on the
strange and axial ones, which are measured via polarized electron scattering. 
Due to the large number of form factors involved in the process, the 
scattering on a single proton is not sufficient to disentangle the 
interesting quantities and heavy targets are needed to yield complementary
information: hence the necessity of controlling the nuclear dynamics in such 
kinematical conditions.
Besides this practical motivation there is a more interesting one from the
nuclear physics' point of view: the scattering of polarized electrons off
complex targets can shed light on some nuclear correlations 
which are not accessible by unpolarized electrons, as will be shown in 
Section~3.

The traditional approaches to wave functions and operators, used in most 
calculations to describe the high-energy regime, involve
leading-order expansions of the electroweak currents:
such an approach is highly constrained to work only at relatively low 
energies and momenta, so that an exact treatment of relativistic effects
is required.

There are two simple general principles which ought to be respected by any 
consistent treatment of the nuclear problem, independently of the details 
of the nucleon-nucleon interaction and of the theoretical framework adopted to 
deal with nuclear correlations (Hartree-Fock, RPA, Brueckner-Hartree-Fock, etc):
\begin{enumerate}
\item Lorentz covariance: the nuclear current must transform as a 
four-vector under a Lorentz boost;
\item Gauge invariance: the nuclear current must be conserved.
\end{enumerate}

The simplest model in which the above fundamental requirements
can be accomplished is the relativistic Fermi gas model (RFG), 
{\em i.e.,} a system of nucleons moving freely inside the nucleus with
relativistic kinematics.
When focus is placed on the quasielastic region, where high-energy 
knockout of nucleons is kinematically favored, this model, 
while undoubtedly too simple to encompass the aspects of nuclear dynamics, is
nevertheless a convenient place to start in such explorations. 

In the quasielastic regime it is reasonable to expect pions to play a role that
differs from the dynamics typically occurring near the Fermi surface,
where one expects other mesons ($\sigma$ and $\omega$ in particular)
to dominate. For quasielastic scattering the residual interaction of
relevance is principally that between a low-energy hole and a very
high-energy particle, and for this the pion is expected to play an
important role. Accordingly, as the next step after the basic
RFG of non-interacting nucleons,
the most important ingredient of the quasi-elastic 
nuclear responses is believed to be the one-pion exchange (OPE) potential.

In this context a consistent first-order operator, embodying all Feynman
diagrams built out of nucleons and pions with one exchanged pion and
one photon attached to all the possible lines can be set up to represent
the two-body current.  Importantly, the latter is gauge 
invariant~\cite{Ama02}.

This fully-relativistic operator includes both the meson-exchange
currents (MEC) and the so-called correlation currents.  
The latter are often not included in
model calculations because they give rise to contributions
already accounted for in the initial and final nuclear wave functions.  
However, the present approach
is based on an uncorrelated relativistic Fermi gas whose states are
Slater determinants built out of (Dirac) plane waves.  
Within a perturbative scheme one is free to consider the one-pion correlation
contributions to the responses as arising either explicitly in the
wave functions or from an appropriate current operator acting on
unperturbed states: here the choice will be the latter.  Clearly, should
it be possible to sum up the whole perturbative expansion, then the
results obtained starting with the true ``correlated'' wave function
would be exactly recovered.

In what follows I shall illustrate the application of the pionic model
to the unpolarized (Section~2) and polarized (Section~3) quasi-elastic
electron scattering, introduce a ``semi-relativistic'' expansion which
allows to ``mimick'' relativity at any value of energy and momentum
transfers (Section~4) and briefly illustrate the scaling and ``superscaling''
behavior of the model (Section~5).

\section{Parity-conserving electron scattering}

The formalism for the unpolarized, inclusive (e,e$'$) process 
is developed in great detail in a variety of papers (see,
e.g., Refs.~\cite{Don02,Mus94,Bar01}).
Hence here only the basic formulae are reported and particular emphasis
is placed on the  above mentioned Lorentz- and gauge-invariance issues.

In the extreme relativistic limit (ERL), in which the incident electron energy
$\varepsilon \gg m_e$, the cross cross section reads
\begin{equation}
\frac{d\sigma}{d\Omega'_e d\omega}=
\frac{2\alpha^2}{Q^4}\left(\frac{\varepsilon'}{\varepsilon}\right)
\eta_{\mu\nu}W^{\mu\nu}=
\sigma_M\left[v_L R^L(q,\omega) + v_T R^T(q,\omega)\right] \ ,
\label{eq_1}
\end{equation}
where $\Omega'_e=(\theta_e,\Phi_e)$ is the scattered electron solid angle, 
$\alpha$ the fine structure constant, $Q_\mu=(\omega,\nq)$ the
transferred four-momentum, 
$\eta_{\mu\nu}$ and $W^{\mu\nu}$ the leptonic and hadronic tensor,
respectively, $\sigma_M$ the Mott cross section and
$v_{L,T}$ are kinematical factors 
(defined, for example, in~\cite{Don92,Alv01}).
The longitudinal and transverse (with respect to the momentum
transfer $\nq$, which fixes the direction of the $z$-axis) 
response functions $R^L$ and $R^T$ are constructed
as components of the hadronic tensor $W^{\mu\nu}$:
\begin{eqnarray}
R^L(q,\omega)&=&\left(\frac{q^2}{Q^2}\right)^2\left[
W^{00}-\frac{\omega}{q}(W^{03}+W^{30})+\frac{\omega^2}{q^2}W^{33}
\right]\label{eq2a} \\
R^T(q,\omega)&=&W^{11}+W^{22} \ .
\label{eq2b}
\end{eqnarray}
  
If gauge invariance is fulfilled,
implying that $W^{03}=W^{30}=(\omega/q)W^{00}=(q/\omega)W^{33}$, 
then $R^L$ is simply the time component of the
hadronic tensor, namely $W^{00}$.  
Hence $R^L$ and $R^T$ are determined by the nuclear charge and current 
distributions, respectively, and they embody 
the entire dependence upon the nuclear structure.

Working within the framework of the RFG model one can then construct
the electromagnetic currents accounting for the effects 
introduced by pions in first-order perturbation theory (one-pion exchange).

\begin{figure}
\vskip -3.cm
\includegraphics[width=13.5cm]{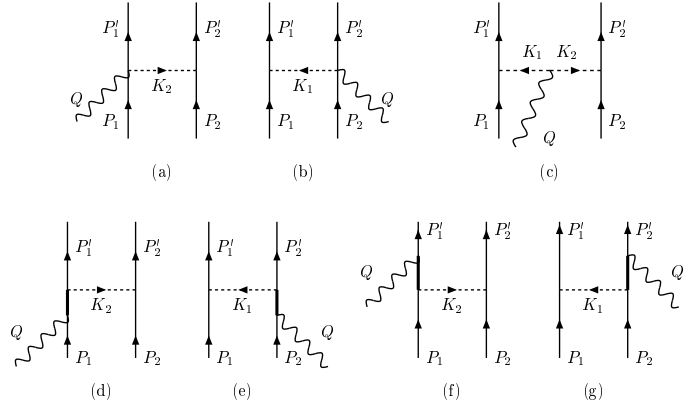} 
\vskip -10cm
\caption{Feynman diagrams contributing to the two-body current 
with one pion-exchange. The wide line in the correlation diagrams
(d)--(g) means a fully-relativistic Dirac propagator for the nucleon.}
\end{figure}

The linked, two-body Feynman diagrams that contribute to electron
scattering with one pion-exchange are shown in Fig.~1. The first three
correspond to the usual meson-exchange currents (MEC): diagrams (a),
(b) refer to the ``seagull'' current, diagram (c) to the
``pion-in-flight'' current. The four diagrams (d)--(g) represent the
so-called correlation current and are usually not treated as genuine
MEC, but as correlation corrections to the nuclear wave function.
However, again note that the present approach puts all correlation effects
in the current operator and uses an uncorrelated wave function for the
initial and final nuclear states.

The general relativistic expressions for the seagull ($s$),
pion-in-flight ($p$) and correlation ($cor$) current matrix elements
of Fig.~1 are (isospin summations are understood)
\begin{eqnarray}
\label{eq6}
&&j^{\mu}_{s}
= \frac{f^2}{m_\pi^2}
             i\epsilon_{3ab}
             \overline{u}(\np'_1)\tau_a\gamma_5\Kbar_1 u(\np_1)
             \frac{F_1^V}{K_1^2-m_\pi^2}
             \overline{u}(\np'_2)\tau_b\gamma_5\gamma^{\mu}u(\np_2)
             + (1 \leftrightarrow 2) 
\\
\label{eq7}
&&j^{\mu}_{p}
=
\frac{f^2}{m_\pi^2}
             i\epsilon_{3ab}
             \frac{F_\pi(K_1-K_2)^\mu}{(K_1^2-m_\pi^2)(K_2^2-m_\pi^2)}
     \overline{u}(\np'_1)\tau_a\gamma_5\Kbar_1 u(\np_1)
             \overline{u}(\np'_2)\tau_b\gamma_5\Kbar_2 u(\np_2)
\\
\label{eq8}
&&
j^{\mu}_{cor}
=           \frac{f^2}{m_\pi^2}
              \overline{u}(\np'_1)\tau_a\gamma_5\Kbar_1 u(\np_1)
              \frac{1}{K_1^2-m_\pi^2} \\
&& 
    \times \overline{u}(\np'_2)
    \left[    \tau_a\gamma_5\Kbar_1 
              S_F(P_2+Q)\Gamma^\mu(Q)
            + \Gamma^\mu(Q)S_F(P'_2-Q)
              \tau_a\gamma_5\Kbar_1
    \right]u(\np_2) 
+ (1\leftrightarrow2) \ . 
\nonumber
\end{eqnarray}
In the above, 
$K_1$, $K_2$ are the four-momenta given to the nucleons 1,
2 by the exchanged pion, and they  
are defined in Fig.~1, while $F_1^V$ and $F_\pi$ are the electromagnetic
isovector nucleon and pion form factors, respectively.  Furthermore,
$u(\np)$ is the free Dirac spinor of a nucleon carrying momentum $\np$,
$S_F(P)$ the nucleon propagator and 
$\Gamma^\mu(Q)=F_1\gamma^\mu+\frac{i}{2m}F_2\sigma^{\mu\nu}Q_\nu$ the
electromagnetic nucleon vertex
(the Galster parametrization is used for the Dirac and Pauli form factors
$F_1$ and $F_2$).

A crucial point to be stressed is that the sum of the relativistic
seagull, pion-in-flight and correlation currents satisfies current
conservation, 
\begin{equation} 
Q_{\mu}(j^{\mu}_{s}+j^{\mu}_{p}+j^{\mu}_{cor})=0\ ,
\label{eq23}
\end{equation}
provided $F_\pi=F_1^V$.
It is also possible~\cite{Mat89,Gro87} to use different phenomenological
electromagnetic form factors for the nucleon and pion --- even
introducing phenomenological form factors at the strong pion-nucleon
vertices --- without violating current conservation, 
by appropriate modification in the currents through the
generalized Ward-Takahashi identity.

Since the interest here is focussed in the one-particle emission induced
by the two-body currents introduced above, one needs now to evaluate
the matrix element of the above two-body operator between the Fermi gas 
ground state and a 1p-1h excitations, which are the dominant modes in the 
quasielastic regime:
\begin{eqnarray}
&&\langle ph^{-1}|\hat{j}^{\mu}(Q)|F\rangle 
=\sum_{k<F} 
\left[ \langle pk |\hat{j}^{\mu}(Q)|hk\rangle
      -\langle pk |\hat{j}^{\mu}(Q)|kh\rangle
\right] \ ,
\label{jmuph}
\end{eqnarray}
where the summation runs over all occupied levels in the ground state,
and thus includes a sum over spin ($s_k$) and isospin ($t_k$) and an
integral over the momentum $\nk$.

The first and second terms in eq.~(\ref{jmuph}) represent the direct
and exchange contribution to the matrix element, respectively.  It can
be easily verified that in
spin-isospin saturated systems the direct term vanishes for the
currents (\ref{eq6}-\ref{eq8}) upon summation over the occupied
states.  Hence only the exchange term contributes to the p-h matrix
elements. 

When inserted in (\ref{jmuph}) the correlation current (\ref{eq8})
gives rise to two kinds of diagrams, called ``vertex corrections'' (VC)
and ``self-energy'' (SE), respectively~\cite{Ama02a,Ama02}.
The self-energy current deserves some special comment, as it diverges:
in fact it corresponds to a SE insertion on an external line, which,
according to field theory, should not be included in a perturbative expansion.
One should then apply a renormalization procedure to dress the external lines
by summing up the entire perturbative series of self-energy
insertions.  In the nuclear context this procedure leads to the
relativistic Hartree-Fock approach.
A renormalized self-energy current corresponding to one-pion-exchange
is constructed in Ref.~\cite{Ama02} by renormalizing spinors and energies and
by expanding the resulting in-medium one-body current to first order
in the square of the pion-nucleon coupling constant.

Once this is done, the electromagnetic inclusive response
functions for one-particle emission reactions can be evaluated 
within the RFG model.
The hadronic tensor that
arises from the interference of the single-nucleon, one-body (OB) current,
$j^\mu_{OB}(\np,\nh)=\overline{u}({\bf p}) \Gamma^\mu u({\bf h})$,
with the one-pion-exchange current,
$j_p^{\mu}+j_s^{\mu}+j_{cor}^{\mu}$,
is for the RFG model with $Z=N$:
\begin{equation}
\ W^{\mu\nu}
=\frac{3Z}{8\pi k_F^3q} \sum_{a=p,s,cor}
\int_{h_0}^{k_F} h dh (\omega+E_{\nh}) \int_0^{2\pi}d\phi_h
\sum_{s_p,s_h} 
\frac{m^2}{E_\np E_\nh} 2{\rm Re}\, 
\left[j^\mu_{OB}(\np,\nh)^*
      j^\nu_a(\np,\nh)
\right] 
\ . \label{eq38} 
\end{equation}

\begin{figure}
\vskip - 2cm
\includegraphics[width=13.5cm]{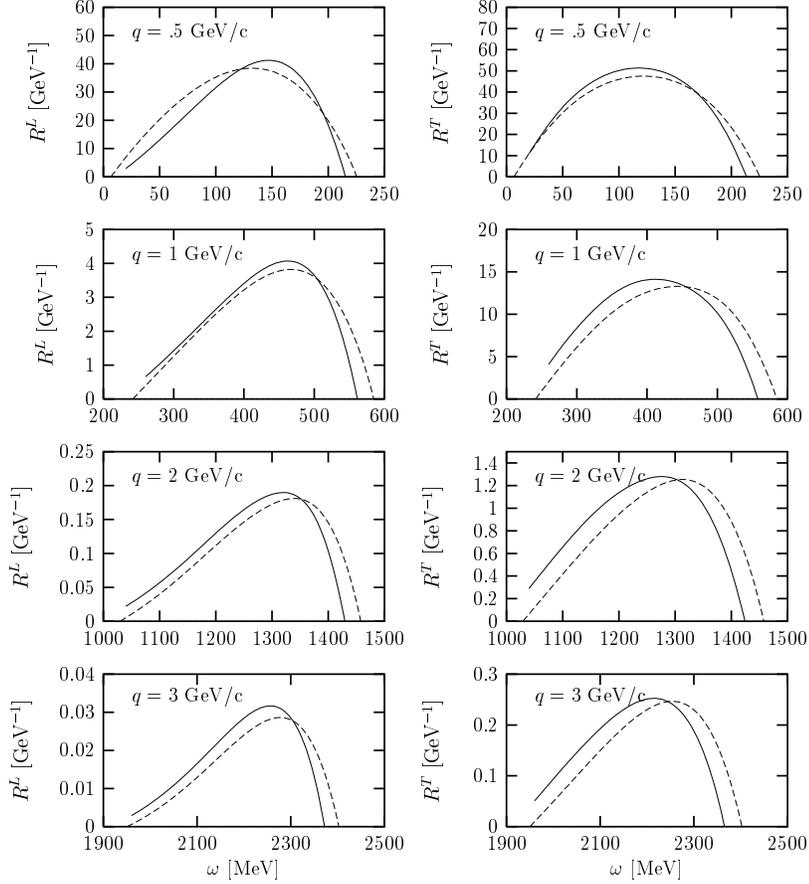}  
\vskip -4.cm
\caption{
Longitudinal (left panels) and transverse (right panels) 
response functions versus $\omega$
including all first-order contributions (solid) compared with the free
result (dashed). The nucleus is
$^{40}$Ca, corresponding to a Fermi momentum $k_F=237$ MeV/c.}
\end{figure}

The numerical results obtained for the 
longitudinal and transverse quasielastic response functions in the
1p-1h sector are shown in Fig.~2.
The calculation refers to $Z=N=20$ and $k_F=237$ MeV/c, 
which is representative of nuclei in the vicinity of $^{40}$Ca.
The total responses are displayed in first
order of perturbation theory and compared with the zeroth-order
ones (free responses) for several momentum transfers.  
In assessing the impact of the global two-body current contribution to the
responses it should be noted that:
\begin{enumerate}
\item the overall effect of the two-body currents appears
sufficiently modest to justify {\em a posteriori} a first-order treatment,
their relative contribution ranging from
$\sim$5 to $\sim$15\% depending upon the kinematics;  
\item the softening at large $q$ appears to be common to both L and T channels,
whereas at low $q$ the longitudinal response displays a hardening that
is absent in the transverse one;
\item the two-body correlation contribution is nearly vanishing 
at the peak of the free responses and is roughly symmetrical about the
quasielastic peak, implying that their impact on the Coulomb sum rule should 
be very small.
\end{enumerate}

\section{Parity-violating electron scattering}

In this section the parity-violating (PV) effects arising
from the weak interaction between the electron and the nucleus
are addressed. Such
effects, which are negligible in unpolarized electron processes, can
be brought to evidence by measuring the asymmetry associated with
longitudinally polarized electrons having opposite helicities.
In this case the purely electromagnetic cross sections cancel out and
one is left with the interference between the electromagnetic and
neutral weak currents, corresponding to the exchange of a photon and a
$Z^0$, respectively.

An important motivation of parity-violating experiments 
is the measurement of the
single-nucleon form factors, in particular the strange and axial ones:
for this reason most experiments are presently being carried out on
light nuclei, where the uncertainties associated with the nuclear
model are minimized. However, other motivations exist for such studies:
specifically, as anticipated in the introduction, the PV response functions
display a different sensitivity to nuclear correlations compared with
the parity-conserving ones: hence they could not only shed light on
the part of the problem concerned with nucleon (and meson) structure,
but also are being used as a test of nuclear models. 
Indeed in~\cite{Alb93,Bar94,Bar96} a
semi-relativistic analysis of the PV responses has been presented,
showing the dominance of pionic correlations in the longitudinal
channel. In~\cite{Ama02} a fully-relativistic calculation has been
performed which confirms the above findings, and extend them to higher 
values of the momentum transfer.

In terms of nuclear response functions the asymmetry reads
\be
{\cal A} \equiv
\frac{d\sigma^+-d\sigma^-}{d\sigma^++d\sigma^-}
= {\cal A}_0 \frac{v_L R^L_{AV}+v_T R^T_{AV}+v_{T'}R^{T'}_{VA}}
{v_L R^L+v_T R^T}\ ,
\label{Asym1}
\ee
where ${\cal A}_0 = \frac{G |Q^2|}{2\sqrt{2}\pi\alpha}$, 
$G$ being the Fermi constant,
and $v_{L,T,T'}$ are leptonic kinematical factors (see 
Refs.~\cite{Don92,Alv01}).
 
The PV response functions are linked 
to the interference hadronic tensor $\tilde W_{\mu\nu}$ by the following 
relations:

\begin{eqnarray}
R^L_{AV}(q,\omega)
&=&
a_A \left(\frac{q^2}{Q^2}\right)^2\left[
\tilde W^{00}-\frac{\omega}{q}(\tilde W^{03}+\tilde W^{30})+
\frac{\omega^2}{q^2}\tilde W^{33}
\right]\label{eq2PVa} \\
R^T_{AV}(q,\omega)
&=&
a_A\left(\tilde W^{11}+\tilde W^{22}\right) \\
R^{T'}_{VA}(q,\omega)
&=&
-i a_V\left(\tilde W^{12}-\tilde W^{21}\right) \ ,
\label{eq2PVb}
\end{eqnarray}
where $a_A=-1$ and $a_V=4\sin^2\theta_W-1$.
The subscript $AV$ in the PV responses denotes interferences of
axial-vector leptonic currents with vector hadronic currents, and the
reverse for the subscript $VA$.

Within the context of the RFG model the interference hadronic tensor is
\begin{equation}
\ \tilde W^{\mu\nu}
=\frac{3Z}{8\pi k_F^3q}
\int_{h_0}^{k_F} h dh (\omega+E_{\nh}) 
\int_0^{2\pi}d\phi_h \sum_{s_p,s_h}
\frac{m^2}{E_{\np}E_{\nh}}
2{\rm Re}\,
\left[j^\mu_{em}(\np,\nh)^* j^{\nu}_{wn}(\np,\nh)
\right] \ , 
\label{tilwrfg}
\end{equation}
where the electromagnetic current $j^\mu_{em}$ includes both the
single nucleon one-body and the two-body
currents discussed in the previous section, {\em  i.e.}
$j^\mu_{em}=j^\mu_{OB}+j^\mu_{MEC}+j^\mu_{cor}$.
The weak neutral current $j^{\nu}_{wn}$ is instead purely one-body,
since the direct coupling of a $Z^0$ to the pion is neglected.

As for the parity-conserving sector, the one-body contribution to the 
three PV responses (\ref{eq2PVa}-\ref{eq2PVb}) can be evaluated
analytically in RFG (see, for example~\cite{Don92} for the explicit 
expressions of the response functions), whereas
the two-body contributions  involve
multidimensional integrals, to be numerically evaluated.

\begin{figure}
\vskip - 1.5cm
\includegraphics[width=13.5cm]{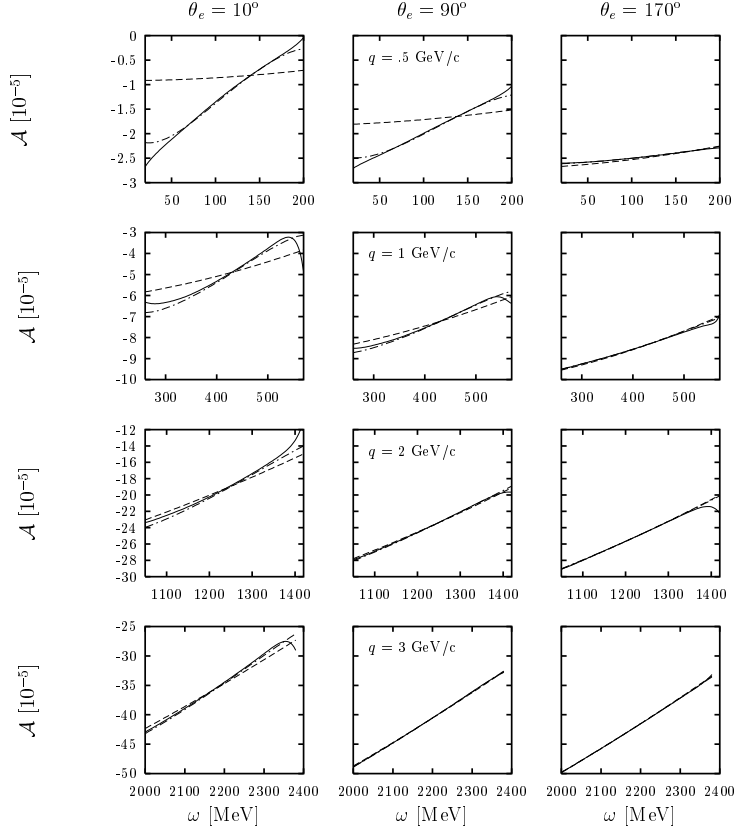}  
\vskip -5.cm
\caption{The PV asymmetry displayed versus $\omega$ for various values of 
the momentum transfer $q$ and the scattering angle $\theta_e$.
Dashed: one-body; dot-dashed: one-body+MEC+VC; solid: total.
}
\end{figure}

In Fig.~3 the effect of the pionic physics
on the asymmetry in eq.~(\ref{Asym1}) is shown by displaying ${\cal A}$
versus $\omega$  
at various values of the momentum transfer $q$ and of the electron scattering
angle $\theta_e$.
The curves represent the free RFG (dashed), and the RFG including MEC and
VC (dot-dashed) or the MEC, VC and SE (solid) contributions.
It appears that the pionic correlations are mostly felt at low values of
$\theta_e$ and $q$.
Indeed a careful analysis, carried out in Ref.~(\cite{Ama02a}), 
shows that the main effect on the asymmetry arises from the vertex
corrections, which dominate at low $q$, in the longitudinal response function.
The latter is enhanced at low $\theta_e$ by the kinematical factor $v_L$:
hence the large modification of the asymmetry at low angles and momentum 
transfers.
On the other hand at large scattering angles
the asymmetry is totally insensitive to  pions,
because the effect of the SE (which gives the main contribution)
cancels between the PV and PC responses appearing in the numerator and
denominator of eq.~(\ref{Asym1}).

The conclusion can be drawn that the extraction 
(at large electron angles) of the
axial nucleonic form factor $G_A$ is almost independent of the nuclear
model. On the contrary at small angles PV experiments can measure the
strange electric content of the nucleon only if a good control of the
nuclear dynamics is achieved, since the isospin correlations give very
large effects.  Conversely, interesting insight into the latter can in
principle be gained here.  The results show that only at very large
momentum transfer does the forward-angle asymmetry become insensitive
to pionic correlations and hence suitable for assessing the
strangeness content of the nucleon.

\section{The $\eta_F$-expansion}

\begin{figure}[hb]
\vskip -2.cm
\includegraphics[width=13.5cm]{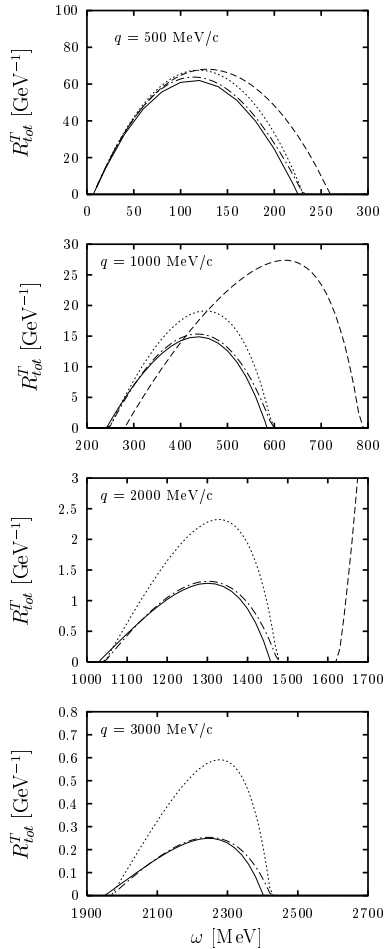}  
\vskip -4.cm
\caption{Total transverse response function of
$^{40}$Ca including MEC for several values of the momentum transfer,
and for $k_F=237$ MeV/c.  Solid: exact relativistic results.  The rest
of the curves have been computed using the non-relativistic Fermi gas
model, with or without relativistic corrections.  Dashed: traditional
non-relativistic results.  Dotted: including relativistic kinematics
in the non-relativistic calculations.  Dot-dashed: including in
addition the new expansion of the OB+MEC currents.  The relativistic
calculations include a dynamical propagator and $\pi N$ form factor,
while the non-relativistic calculations do not include these
corrections.}
\end{figure}

In the non-relativistic reductions commonly used in treating the effects of
two-body pionic currents in electron scattering 
reactions, e.g.\cite{Van81,Alb90,Ama94a},
not only non-relativistic wave functions have been used, but also 
non-relativistic current operators derived from a direct Pauli reduction 
have been considered. Although these approximations are sufficient
in the low-energy regime, they badly fail to hold at high momentum
transfer.

On the other hand the exact relativistic calculation outlined in Section~2 is
limited, up to now, to the simple pion-exchange potential and to first
order perturbation theory.
These two assumptions are not too crude in the
quasi-elastic region, but they cannot obviously be applied to 
different kinematical regimes, where higher orders in the perturbative 
series and different correlations (in particular short-range correlations) 
are known to be far from negligible.

In order to implement effects of relativity into more realistic
nuclear models, new expressions of the one- and two-body
electromagnetic currents have been 
suggested in~\cite{Jes98,Ama96a}, as well as the $N\to\Delta$
electromagnetic current~\cite{Ama99a}: here the current is derived as a
non-relativistic expansion in terms of the dimensionless parameter
$\eta\equiv p/m$, $p$ being the three-momentum of the struck
nucleon. 
Since the latter is limited by the Fermi momentum, the
procedure amounts to an expansion in the parameter 
$\eta_F=k_F/m\simeq 1/4$. 
It should be stressed that such expansion holds valid for any energy
and momentum transfer and can therefore be applied even in extreme
kinematical conditions.

Although the $\eta_F$-expansion is not needed in the present context,
where the calculations are performed without any non-relativistic 
approximation, it has the merit of yielding recipes to include
relativistic corrections in the non-relativistic currents through 
the simple kinematical factors. For example 
the relation between the non-relativistic ($NR$) and the new 
``semi-relativistic'' ($SR$) pion-in-flight current so 
obtained is found to be~\cite{Ama98a}:
\be
j^\mu_{p,SR} = \frac{1}{\sqrt{1+\tau}} j^\mu_{p,NR}\ ,
\ee
where $\tau=|Q^2|/(4m^2)$.
Similar relations can be derived for all the components of the 
one- and two-body nuclear current.
By means of this expansion one could in principle extend the validity of
existing sophisticated, but non-relativistic, calculations to regimes where 
they are not at present applicable.

To appreciate the quality of the method in Fig.~4 the
total transverse response, including OB+MEC operators, is reported for 
$q=500,$ 1000, 2000, and 3000 MeV/c. The solid lines are the exact
relativistic result, the dashed lines represent the traditional
non-relativistic results, which together with relativistic kinematics
give the dotted lines. Finally, the dot-dashed lines correspond to the 
results using the semi-relativized OB+MEC currents. 
Whereas the non-relativistic approach is clearly failing already at 
moderate values of $q$ (of the order $500~MeV/c$),
the agreement between the exact model and the $\eta_F$-expanded one
is quite good even for very high $q$:
these currents are therefore very
appropriate and easy to implement in already existing non-relativistic
models for the electromagnetic reactions. 

\section{Scaling}

Before concluding I will shortly illustrate the behavior of pionic 
correlations in the quasi-elastic peak with respect to scaling and
superscaling.
A comprehensive introduction to the concept of scaling can be found in 
Refs.~\cite{Don99a,Don99b,Mai02,Don02}. 
Here I only remind the basic definitions:
\begin{enumerate}
\item
Scaling of I kind occurs if the ratio $F^{L,T}(q,\omega)$ between the
nuclear response functions 
$R^{L,T}(q,\omega)$ and the relevant single-nucleon electromagnetic
factors becomes, for high values of $q$,
function of one single variable, the scaling variable.
Several different scaling variables exist in the literature, all of them
coalescing into one - or being simply related to each other - for high
enough momentum transfers. In the quasi-elastic peak region the 
natural scaling variable turns out to be~\cite{Alb93,Bar98}
\be
\psi=\pm\sqrt{\frac{T_0}{T_F}}\ ,
\label{psi}
\ee
being 
\be
T_0=\frac{1}{2}\left(q\sqrt{1+\frac{4m^2}{|Q^2|}}-\omega\right)
\ee
the minimum kinetic energy required to a nucleon to take part in the
process and $T_F$ the Fermi kinetic energy. The $+(-)$ sign in 
(\ref{psi}) refers to the right (left) of the quasi-elastic peak.
The analysis of the world data~\cite{Don99a,Don99b,Mai02} 
shows that scaling of I kind is reasonably
good for $\psi<0$ and badly violated for $\psi>0$.
\item
Scaling of II kind consists in the independence of the functions
$f^{L,T}=k_F\times F^{L,T}$ on the specific nucleus, namely on the Fermi 
momentum. The analysis of the existing data points to an excellent 
fulfillment of this scaling in the region $\psi<0$ and to a not very 
dramatic breaking of it for $\psi>0$.
When the two kinds of scaling occur the responses are said to ``superscale''.
\end{enumerate}

The relativistic Fermi gas model fulfills both kinds of scaling by 
construction.
The observed superscaling behavior of the experimental 
data~\cite{Don99a,Mai02} offers a clear 
constraint to the nuclear correlations and can be used as a test of the 
reliability of the model.
It is then natural to ask whether or not the present pionic model 
superscales.

In~\cite{Ama02a} the evolution with $q$ of
the MEC in the transverse channel has been explored in detail:
it has been proven that the relative
contribution of the MEC to $R^T$ decreases with $q$, but does not vanish for
large values of $q$. In fact, it decreases
in going from 0.5 to 1 GeV/c, but then it rapidly saturates at or
slightly above $q$=1 GeV/c, where its value stabilizes, typically
around 10$\%$. Thus, one can conclude that at momentum transfers above
1 GeV/c scaling of the first kind is satisfied for the MEC
contributions considered in this work.  Moreover, it is found that
for high $q$ the MEC almost vanish for $\omega$ in the vicinity of the QEP.

The evolution with $q$ of the correlation current in the longitudinal and
transverse channels has also been discussed at length 
in~\cite{Ama02a}. The basic findings are that:
1) the VC do not saturate quite as rapidly as the MEC, although their
behavior is rather similar and saturation again occurs somewhere
above $q=1$--1.5 GeV/c:  thus, again,
scaling of the first kind is
achieved at high momentum transfers for these contributions.
Moreover, similarly to the MEC case, for high $q$ the VC almost
vanish around the QEP. 
2) A somehow different behavior is observed in the self-energy, which, due to
a delicate cancellation between the particle and hole dressings,
grows - instead of decreasing - with $q$ in the
range $q$=0.5--2 GeV/c, and then stabilizes typically at about 30-40\% of
the free response to the left of the QEP, thus inducing an important
softening of the longitudinal and transverse responses.  

As far as scaling of second kind is concerned,
the two-body current contribution to the response functions 
is found to {\em grow} with $k_F$, in contrast with
the free response which decreases as $k_F^{-1}$.  More specifically
the two-body processes violate the second-kind scaling by
roughly three powers of $k_F$.  
This effect is a rapid function of the
Fermi momentum (or equivalently, of the density): for example, if one
considers the cases $^2$H/$^4$He/heavy nuclei with Fermi momenta of
approximately 55/200/260 MeV/c, respectively, then the 1p-1h MEC
contributions amount to 0.1/5/10\% of the total transverse response,
respectively (normalizing to 10\% for the heavy nucleus case).

In summary, in the present pionic model scaling of the first kind is 
achieved at momentum transfers somewhat below 2 GeV/c, whereas scaling of
second kind is badly violated. 
A similar trend is expected to be followed by the contributions of the
heavier mesons, which have been neglected in the present approach: in order 
to agree with the experimental data, the strength of these 
contributions cannot be too disruptive.

\section{Conclusions}

The impact of pionic correlations and meson-exchange currents
on the nuclear electromagnetic response functions,
calculated in a fully relativistic context which allows to respect exactly
gauge invariance, is found to be modest in the quasi-elastic region.
On the other hand the parity-violating asymmetry displays a strong
sensitivity to such correlations if the scattering angle and momentum
transfers are small. 
The analysis of the corrections induced by the pion with respect to
``superscaling'' shows that, whereas they do not disrupt the I-kind 
scaling of the relativistic Fermi gas, they badly violate scaling of
II-kind.

Of course the pion is only one ingredient of the NN force, which, due
to its long-range nature, is thought to give the dominant contribution 
to the responses in the quasi-elastic peak region. In order to deal with
different observables and kinematical regions,
where the short-range physics is known to play an important role, 
the present model should
be extended to include heavier mesons exchange and currents. A first effort 
in this direction is performed in Ref.~\cite{Ama02c}, where the 
modification of the momentum distribution due to the full Bonn potential
is analyzed.

\acknowledgments
The work presented was carried out in collaboration with J.E.~Amaro,
J.A.~Caballero, T.W.~Donnelly and A.~Molinari. I would like to thank 
J.E.~Amaro and J.A.~Caballero for carefully reading the manuscript and to 
acknowledge financial support from MEC (Spain) for a sabbatical stay at 
University of Sevilla (SAB2001-0025).

\end{document}